# Study of the effect of Sharpness on Blind Video Quality Assessment

Anantha Prabhu[1], David Pratap[2], Narayana Darapeni[3] and Anwesh P R[2,*]

[1] PES University, Bangalore, Karnataka, 560050, India
[2] Great Learning, Hyderabad, Telangana, 500089, India
[3] Northwestern University, Evanston, IL 60208, United States

## Abstract

INTRODUCTION: Video Quality Assessment (VQA) is one of the important areas of study in this modern era, where video is a crucial component of communication with applications in every field. Rapid technology developments in mobile technology enabled anyone to create videos resulting in a varied range of video quality scenarios.
OBJECTIVES: Though VQA was present for some time with the classical metrices like SSIM and PSNR, the advent of machine learning has brought in new techniques of VQAs which are built upon Convolutional Neural Networks (CNNs) or Deep Neural Networks (DNNs).
METHODS: Over the past years various research studies such as the BVQA which performed video quality assessment of nature-based videos using DNNs exposed the powerful capabilities of machine learning algorithms. BVQA using DNNs explored human visual system effects such as content dependency and time related factors normally known as temporal effects.
RESULTS: This study explores the sharpness effect on models like BVQA. Sharpness is the measure of the clarity and details of the video image. Sharpness typically involves analyzing the edges and contrast of the image to determine the overall level of detail and sharpness.
CONCLUSION: This study uses the existing video quality databases such as CVD2014. A comparative study of the various machine learning parameters such as SRCC and PLCC during the training and testing are presented along with the conclusion.

**Keywords:** VQA - Video Quality Assessment, BVQA – Blind Video Quality Assessment, SSIM - Structural Similarity Index Measure, PSNR - Peak Signal to Noise Ratio, CNN - Convolutional Neural Network, MOS – Mean Opinion Score, ResNet50, ResNet152, SRCC - Spearman's rank correlation coefficient, PLCC- Pearson Linear Correlation Coefficient.

## 1. Introduction

Video Quality Assessment (VQA) is one of the important areas of study in this modern era, where video is a crucial component of communication with applications in every field. Rapid technology developments in mobile technology enabled anyone to create videos resulting in a varied range of video quality scenarios. Though VQA was present for some time with the classical metrices like SSIM and PSNR, the advent of machine learning has brought in new techniques of VQAs which are built upon Convolutional Neural Networks (CNNs) or Deep Neural Networks (DNNs).

VQA dates back to the earliest days of digital video, when the primary issue was ensuring that the video was encoded and sent without error. As quality of digital video improved and use of video in numerous applications grew, the focus of VQA switched to evaluating the users end-perception of the video's quality.

Throughout the years, a few distinct VQA methodologies have been created, including:
A video analysis method that uses mathematical algorithms to determine the film's quality. This comprises signal-to-noise ratio (PSNR) [1] and structural similarity index (SSIM) [2]. Subjective VQA [3]: which entails asking human viewers to rate the video's quality. Typically, this is accomplished through the use of Mean Opinion Score (MOS) assessment. No-reference VQA [4]: which determines the video's quality without reference to a perfect version. Full-reference VQA [5]: a video quality assessment method that compares the video under test to a flawless version and is often used in video compression. Reduced reference VQA [6]: which utilizes a subset of the information from the original version. VQA is crucial in numerous application domains, including video coding, transmission, and storage. In addition, it ensures that the video content is of good quality and that the viewer's experience is not diminished.

Over the past years various research studies such as the BVQA (Blind Video Quality Assessment) [7] which performed assessment of quality of nature-based videos using DNNs exposed the powerful capabilities of machine learning algorithms. BVQA using DNNs explored the human visual system effects such as content dependency and time related factors normally known as temporal effects.

Large set of initial BVQA models are built upon NSS. Natural Scene Statistics (NSS) [8] refers to the statistical properties of natural scenes that are captured by cameras or observed by human vision. The idea behind using NSS for image and video quality assessment is that natural scenes exhibit certain statistical regularities, and any deviations from these regularities could indicate image or video distortions. Classical Blind/Reference less Image Quality Assessment (BVQA) models rely heavily on NSS analysis, and some of the most common NSS features used in these models include natural image statistics, such as image sharpness, contrast, and colorfulness. While these models have shown some success in predicting quality scores, they are not without limitations. For example, they tend to perform poorly when the distortions are not well captured by NSS, and they can also be sensitive to changes in the content of the images or videos being analyzed. Nonetheless, NSS-based BVQA models remain an important area of research and study in the realm of image and video processing.

In the last few years, deep neural networks (DNNs) have become increasingly popular in the field of Blind/Reference less Image Quality Assessment (BVQA), challenging the classical NSS-based models. DNNs have demonstrated remarkable representation learning capabilities and have shown promising results in enhancing computer vision applications, such as classification of images, detection of objects, and semantic segmentation. DNN-based BVQA models have the potential to address some of the limitations of classical models, such as their sensitivity to content changes and the limited generalization capability across different distortion types.

The success of DNN-based BVQA models is due to their ability to learn complex feature maps from the input data and capture higher-order dependencies between the features. The key challenge in training DNNs for BVQA is to obtain a large dataset of images or videos with corresponding subjective quality scores. To this end, several image and video quality assessment datasets have been released, including CVD2014 [10], LIVE-Qualcomm [11], YouTube-UGC [12], and KoNViD-1k [9], to name a few. These datasets contain many distorted images or videos with corresponding subjective quality scores obtained from human observers.

Subjective quality testing is considered the gold standard of video quality assessment as it measures the perception of video quality by humans, who are the ultimate consumers of videos. However, subjective testing is labor-intensive and takes long time duration. This makes it difficult to scale up to large-scale applications. As an viable alternative, objective video quality assessment (VQA) methods have emerged, aimed at predicting the perceived quality of videos automatically.

Objective VQA methods use computational models to assess the video quality based on various visual features, such as spatial and temporal distortions, contrast, and sharpness. Objective VQA has many advantages over subjective testing, including cost-effectiveness, scalability, and faster turnaround time. It can also evaluate the quality of video in real-time applications, like video conferencing and streaming.

Among objective VQA methods, Blind/Reference less Video Quality Assessment (BVQA) has gained significant attention in recent years. BVQA aims to predict video quality without reference to an original, undistorted version of the video.

In this research study, we propose a BVQA approach that we refer to as Blind Video Quality Assessment. This method uses multistage CNNs to arrive at a quality assessment. A special focus area is to access the effects of Sharpness features on the BVQA model.

## 2. Materials and Methods

### 2.1. Dataset

A dataset is a crucial component for developing a blind video quality assessment (BVQA) model. The dataset used for training and evaluation should be diverse and representative to ensure that the results are generalizable. A good BVQA dataset should include various types of videos with different resolutions, frame rates, and content. The dataset could also include videos with various kinds of distortions such as compression artifacts, blur, noise, and blockiness. Furthermore, the dataset should also consider the subjective nature of human perception by including subjective quality ranks obtained from human observers. The inclusion of subjective quality ranks allows for the evaluation of the BVQA model against the human perception of quality of the video. Overall, the selection and preparation of the dataset is crucial for the development and evaluation of accurate BVQA models.

The field of video quality assessment has witnessed significant progress in the last few years, thanks in part to the availability of huge datasets such as CVD2014, KonVid-1k, LiveQualcomm, and YouTube-UGC. These datasets contain thousands of videos and corresponding subjective ratings, allowing researchers to train and test video quality models on a wide variety of content. However, despite the benefits of these datasets, there are limitations to their use. For example, the sheer size of these datasets can make it difficult to process and analyze them using existing computing resources. In addition, some datasets may not be publicly available, making it difficult for researchers to replicate previous work. In the case of our project, we were only able to use a small subset of the CVD2014 dataset due to limitations in our system. While this was a significant limitation, it allowed us to still make progress towards our research goals, and we hope to expand our dataset in future phases of the project.

## 2.2. Data Preparation

The CVD21014 dataset is a good benchmark dataset for assessment of quality of in-the-wild videos. The CVD2014 dataset contains 116 distorted video sequences, which were collected from YouTube and Vimeo and cover an exhaustive range of video content and quality levels. The dataset also includes the corresponding pristine video sequences for each distorted video, leading to a total of 234 video sequences.

The videos in the CVD21014 dataset were encoded using H.264/AVC with a fixed bitrate of 2 Mbps and a resolution of 720p. The sequences were then distorted using five types of commonly encountered video distortions, including compression artifacts, blurring, noise, blockiness, and temporal misalignment.

To arrive at a subjective quality rank for the distorted videos in the CVD21014 dataset, an experiment to assess subjective quality was conducted using the Absolute Category Rating (ACR) method. In this experiment, a group of human observers was presented with pairs of distorted videos and instructed to rate which video had better quality. The ranks were then converted into mean opinion scores (MOS) using the maximum likelihood estimation method.

For the current study, we used a total of 54 videos from the Test1 and Test2 directories. These 54 videos are used for training and validation of the proposed model. MatLab script is created, which will read the CVD2014 video names from the CVD2014 directory. It will also read the MOS scores and details provided as part of the xlsheet from the dataset. It creates the .mat file. This .mat file would be the input file for all the feature extractor modules.

The limited CVD21014 dataset was split into three sets, one each for training, testing, and validation purposes. Specifically, 60% of the videos were used during the training phase, 20% for the validation phase, and 20% for the testing phase. The training data set was used to train the BVQA model, while the validation data set was used to tune the hyperparameters of the model and to avoid overfitting. The testing data set was used to evaluate the model performance on unseen videos.

## 2.3 Models

The overall BVQA model is illustrated in Figure 1. The illustration shows an example of a multistage model.

In the quality aware pre-training stage, the model is trained to learn a representation that is effective for quality assessment by leveraging both quality-aware and quality-agnostic features. Specifically, two pre-training modules are utilized: a video classification module trained on the Kinetics-400 dataset, and a video quality assessment module trained on the TID2013 dataset. The video classification module learns features that are insensitive to quality variations, while the video quality assessment module extracts features that are sensitive to quality of video.

In the motion perception stage, model undergoes fine-tuning to enable quality assessment of videos captured in real-world conditions by incorporating motion information. The motion information is extracted using a simple motion detection algorithm, and the fine-tuning module adapts the pre-trained model to the motion perception features using the CVD2014 dataset. Finally, the quality prediction module predicts the video quality of the input video based on the adapted model using a support vector regression (SVR).

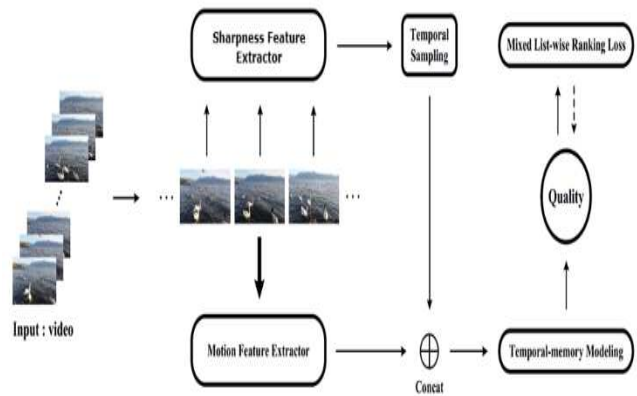

Figure. 1. Modified BVQA Model.

## 2.4 Sharpness Feature Extractor

The Sharpness Feature Extractor is the one which replaces the Spatial Feature extractor module. The sharpness feature ideally depicts the clarity and edge features of an image. The sharpness feature extractor is designed to collect high-level sharpness features from input video frames, which are then used for quality assessment.

The Sharpness feature extractor is based on pretrained resnet18 model. The last two resnet layers in the pretrained

resnet18 model are unfrozen to learn the sharpness features. The Sharpness feature extractor is trained on a TID2013 which is mainly a synthetic database containing 25 original images. Each image is synthetically modified to add various types of noises. The sharpness extractor is trained mainly on those set of distortions which impact the sharpness features of the image. Thus, Sharpness feature extractor model is trained on a set of 900 total images having varying degree of sharpness related distortions.

The output of the sharpness feature extractor is a sharpness tensor of dimension 1024. This tensor represents the high-level sharpness features extracted from the input image frame from the video, which are then fed into the subsequent layers of the proposed model for quality prediction. The use of a high-dimensional feature tensor allows the model to capture fine-grained sharpness details of the input image frame from the video that are relevant to image quality.

In summary, the sharpness feature extractor in the proposed model is a CNN that is pre-trained on specific set of images that have sharpness related distortions from the TID2013 dataset, to extract high-level sharpness features from the TID2013 input images. This sharpness features specific model is then used on the CVD2014 videos to obtain the sharpness features that are further used in the BVQA model for training/validation.

## 2.5 Motion Feature Extractor

The Motion Feature Extractor module uses the SlowFast [14] network, which is a convolutional neural network (CNN) architecture that is devised for action recognition in videos. A SlowFast network is illustrated in Figure 2. The SlowFast network architecture used here, is pre-trained on the Kinetics 400 dataset [13], which contains videos of human actions in different environments and conditions.

The motion feature extractor module is made up of two main components: a SlowFast network for feature map extraction and a pooling layer for feature aggregation. The SlowFast network takes in the input video frames and extracts spatio-temporal features at different temporal resolutions using a combination of slow and fast pathways. The slow pathway collects spatial information at a low temporal resolution, while the fast pathway collects motion information at a high temporal resolution. The output of the SlowFast network is a set of feature maps that represent the spatio-temporal insights into the video. The pooling layer then aggregates the feature maps over the entire video sequence to obtain a single feature vector that represents the quality of video. The pooling layer uses a weighted average of the feature maps, where the weights are learned using a quality-aware pre-training scheme. Various SlowFast network layers are illustrated in Figure *3*.

The motion feature extractor module using the SlowFast network is effective in capturing different types of motion cues that are indicative of video quality, such as blurring, jerkiness, and temporal consistency. The pre-trained SlowFast network on the Kinetics 400 dataset provides a good starting point for capturing complex spatio-temporal features in videos, while quality-aware pre-training scheme enables module to learn the weights that are specific to video quality assessment. Overall, the motion feature extractor module plays a crucial role in the proposed blind VQA system, enabling it to assess video quality without access to the original video.

The motion feature tensor is obtained by feeding the input video frames into the SlowFast network, which extracts spatio-temporal features at different temporal resolutions using a combination of slow and fast pathways. The resultant output of the SlowFast network is a set of feature maps that represent the spatio-temporal information in the video. These feature maps are then aggregated using a pooling layer, which produces a motion feature tensor of dimension 512.

The motion feature tensor contains rich information about the motion characteristics of the video, such as the magnitude and direction of motion vectors, the energy of the motion field, and the spatial and temporal consistency of the motion. This information is used to depict the quality of the video, and the motion feature tensor is fed into a quality prediction module that produces a quality score for the video.

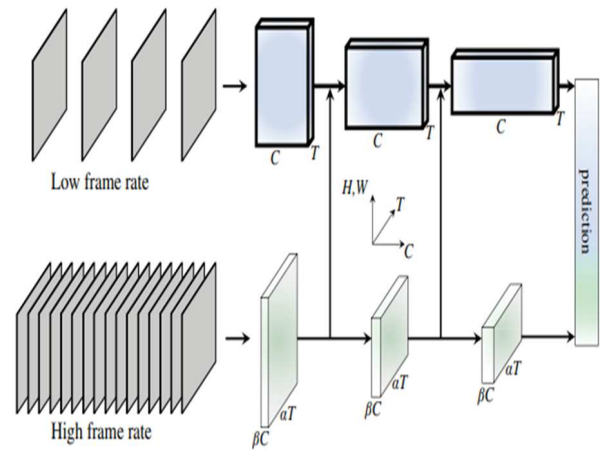

Figure. 2. Slow Fast Network.

| stage | Slow pathway | Fast pathway | output sizes $T \times S^2$ |
|---|---|---|---|
| raw clip | - | - | $64 \times 224^2$ |
| data layer | stride 16, $1^2$ | stride 2, $1^2$ | Slow: $4 \times 224^2$ <br> Fast: $32 \times 224^2$ |
| $conv_1$ | $1 \times 7^2$, 64 <br> stride 1, $2^2$ | $5 \times 7^2$, 8 <br> stride 1, $2^2$ | Slow: $4 \times 112^2$ <br> Fast: $32 \times 112^2$ |
| $pool_1$ | $1 \times 3^2$ max <br> stride 1, $2^2$ | $1 \times 3^2$ max <br> stride 1, $2^2$ | Slow: $4 \times 56^2$ <br> Fast: $32 \times 56^2$ |
| $res_2$ | $\begin{bmatrix} 1 \times 1^2, 64 \\ 1 \times 3^2, 64 \\ 1 \times 1^2, 256 \end{bmatrix} \times 3$ | $\begin{bmatrix} 3 \times 1^2, 8 \\ 1 \times 3^2, 8 \\ 1 \times 1^2, 32 \end{bmatrix} \times 3$ | Slow: $4 \times 56^2$ <br> Fast: $32 \times 56^2$ |
| $res_3$ | $\begin{bmatrix} 1 \times 1^2, 128 \\ 1 \times 3^2, 128 \\ 1 \times 1^2, 512 \end{bmatrix} \times 4$ | $\begin{bmatrix} 3 \times 1^2, 16 \\ 1 \times 3^2, 16 \\ 1 \times 1^2, 64 \end{bmatrix} \times 4$ | Slow: $4 \times 28^2$ <br> Fast: $32 \times 28^2$ |
| $res_4$ | $\begin{bmatrix} 3 \times 1^2, 256 \\ 1 \times 3^2, 256 \\ 1 \times 1^2, 1024 \end{bmatrix} \times 6$ | $\begin{bmatrix} 3 \times 1^2, 32 \\ 1 \times 3^2, 32 \\ 1 \times 1^2, 128 \end{bmatrix} \times 6$ | Slow: $4 \times 14^2$ <br> Fast: $32 \times 14^2$ |
| $res_5$ | $\begin{bmatrix} 3 \times 1^2, 512 \\ 1 \times 3^2, 512 \\ 1 \times 1^2, 2048 \end{bmatrix} \times 3$ | $\begin{bmatrix} 3 \times 1^2, 64 \\ 1 \times 3^2, 64 \\ 1 \times 1^2, 256 \end{bmatrix} \times 3$ | Slow: $4 \times 7^2$ <br> Fast: $32 \times 7^2$ |
| | global average pool, concate, fc | | # classes |

Figure. 3. Slow Fast Network Layer details.

In summary, the motion feature extractor module produces a motion feature tensor of dimension 512 that captures the spatio-temporal features of the input video frames. This tensor contains rich information about the motion characteristics of video, which depicts the quality of video. The motion feature tensor is a key component of the proposed blind VQA system, enabling it to assess video quality without access to the original video.

## 2.6 Fusion layer

As discussed in the previous sections, a ResNet-18 as part of sharpness feature extraction layer used to collect visual sharpness specific features from the input frames of a raw video clip. The sharpness information is aggregated using global average pooling (GAP) and global standard deviation pooling (GSP) to obtain a rich sharpness feature representation of each frame. This results in 1024-dimensional feature vectors for each frame with a temporal length of T.

For the motion feature extraction, the model uses the SlowFast network with default parameters described in [14]. The SlowFast network produces an activation map with a size of {T/2, H/32×W/32, 256} for each video clip z. Similar to the spatial feature extraction, the features are pooled using GAP and GSP, leading to a sequence of 512-dimensional frame-level feature map {vm,t} for each clip with a temporal length of T/2.

To fuse the sharpness features map and motion feature map, the model temporally samples one out of every two frames of the sharpness feature tensor to match the temporal resolution of the motion pipeline. Then, the sharpness and motion feature maps are concatenated along the channel dimension to obtain 1536-dimensional frame-level feature vectors for each clip with a temporal length of T/2. This fused feature tensor is used to produce a quality score for the video using a fully connected layer.

## 2.7 Model Loss Measure

The model employs two loss functions to achieve this goal of consistently estimating the quality of videos as rated by humans: the Pearson Linear Correlation Coefficient (PLCC) loss and the Spearman Rank-order Correlation Coefficient (SRCC) loss.

The PLCC loss helps optimize the model for higher estimation precision by measuring the degree of linear correlation between estimated quality scores and ground truths. The study uses a nonlinear mapping with a 4-parameter logistic function to calculate the PLCC. The SRCC loss aims to boost the prediction monotonicity of the model, which refers to the ability of the model to predict video quality consistently in a ranking order. Unlike the PLCC loss, which is commonly used in video quality assessment, the SRCC loss is not differentiable due to the non-differentiable operations of frequently used order statistics and ranking metrics. To overcome this issue, the authors adopt a differentiable proxy that computes "soft" ranks and is differentiable. The overall loss function of the model is a combination of the PLCC and SRCC losses.

## 2.8 Model Performance Measure

Blind video quality assessment (BVQA) models are used to predict the perceived video quality without the use of reference or original videos. These models can be based on machine learning techniques or deep learning architectures such as convolutional neural networks (CNNs) or recurrent neural networks (RNNs). The selection of BVQA model depends on the complexity of the task, the availability of training data, and the performance requirements. BVQA models can be evaluated using metrics such as Pearson Linear Correlation Coefficient (PLCC) and Spearman Rank Correlation Coefficient (SRCC). These metrics are commonly used in the context of subjective quality assessment to measure the correlation between the estimated quality scores and the mean opinion scores (MOS) provided by human observers. The performance of BVQA models can be improved by using ensemble methods or by combining multiple models. Overall, the development of accurate BVQA models using PLCC and SRCC metrics is important for a wide variety application such as video compression, streaming, and transmission, and ongoing research is focused on improving the accuracy and efficiency of these models.

## 3. Results

The study carried out involved 54 videos of varied length frames each. Each of these videos were associated with a MOS as described in Data preparation section. Training phase used the MOS as one of the inputs to the feature extraction stage.

The study was carried out with different sets of hyperparameters. Batch size, Epochs are the primary hyperparameters which were changed throughout the experiment keeping the learning rate constant.

Training has been carried out for a range of epochs starting with 20 to 100. PLCC and SRCC are documented for each epoch. The whole exercise is repeated with ResNet18 as the Sharpness feature extraction model. The blue line in the SRCC and PLCC graphs indicates the results when the feature extraction was done using the Spatial Feature extractor as in the original model and the red line represents the results when the Sharpness feature extraction was done in the modified model as shown in fig 4.

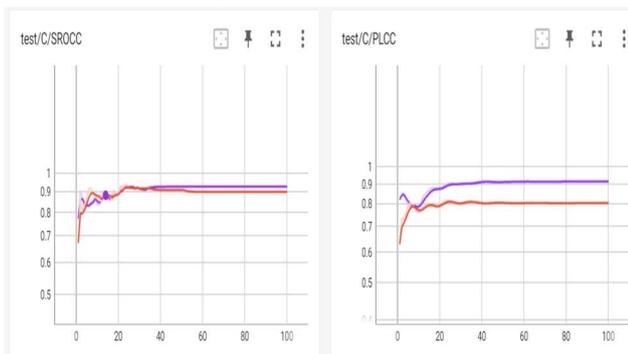

Figure. 4. Modified BVQA SRCC and PLCC

The comparison of the original model against the modified model (with sharpness feature extractor) is as shown in the below table 1

TABLE.1. Criteria Comparison

| Criterion /Model | Comparison | |
|---|---|---|
| | Original BVQA | Modified BVQA |
| SRCC | 0.9 | 0.8954 |
| PLCC | 0.9133 | 0.8788 |

## 4. Discussions and Conclusions

The Phase1 of the study carried out using the limited CVD2014 dataset, and the corresponding findings are presented in the Table 1. The CVD2014 dataset used in this study is a subset of the larger CVD2014 database, consisting of videos with reduced resolution due to resource limitations. The purpose of Phase1 of the study was to arrive at a baseline system that could predict the quality of these reduced videos. The implementation involved training and testing the original BVQA model using the limited CVD2014 dataset and evaluating their performance using standard quality metrics. The models were trained on a combination of objective and subjective quality scores, which were provided with the dataset. The objective scores were calculated using established quality metrics, while the subjective scores were obtained through subjective quality assessments by human observers.

Table 1 presents the results of the implemented models, showing their performance in terms of the Spearman Rank Correlation Coefficient (SRCC) and the Pearson Linear Correlation Coefficient (PLCC). These metrics are commonly used to evaluate the model performance with respect to quality prediction. The results indicate that the implemented models achieved better performance with limited CVD2014 dataset. The best-performing model achieved an SRCC of 0.9 and a PLCC of 0.9133, indicating a strong correlation between the predicted and actual quality scores.

Phase 2 of the study of effects of the Sharpness extractor and the corresponding results as shown in Table 1 which compares the original model with the modified model. As the comparison indicates, the Sharpness feature extractor modified model is coming very close to the results obtained by the original model. But does not exceed the original model results.

This indicates that the Sharpness feature extractor replacement of Spatial features extractor works well with the BVQA.

## 5. Future Possibilities

Video Quality Assessment is a wide topic, Sharpness feature extractor based modified BVQA model came close to the original BVQA model which was using Spatial features extractor. Probable reasons could be the inadequate training of the Sharpness feature extractor, or the limited CVD2014 data set used for training/validation of the models.

As future work, following research activities can be explored to further study the effects of Sharpness features on the BVQA model.

1. Exhaustive training of Sharpness feature extractor with more focused training on the sharpness features to better the Sharpness feature extractor.

2. Use the exhaustive CVD2014 and other available video databases on an enhanced hardware platform to conduct the research.

3. Possible way of stitching a Sharpness feature extractor as the third stage in the original BVQA model instead of replacing the Spatial feature with the Sharpness features.